# A New Quantum Approach to Binary Classification

Krishna sai Mangalarapu [1], Vamsi Krishna [2], Munawar Ali [3]

*Undergrad Students Pursuing Computer science engineering from J B Institute of Engineering and Technology, Hyderabad.*
*Under the supervision of* **Dr. G. Arun Sampaul Thomas (Phd)**

***Abstract :*** *Machine Learning classification models learn the relation between input as features and output as a class in order to predict the class for the new given input. Quantum Mechanics (QM) has already shown its effectiveness in many fields and researchers have proposed several interesting results which cannot be obtained through classical theory. In recent years, researchers have been trying to investigate whether the QM can help to improve the classical machine learning algorithms. It is believed that the theory of QM may also inspire an effective algorithm if it is implemented properly. From this inspiration, we propose the quantum-inspired binary classifier.*

## I. Introduction

Similar to the way conventional computers are made up of bits, quantum computers are made up of quantum bits, or qubits. Just like a bit, a qubit has a state. However, while the state of a bit is a number ( or ), the state of a qubit is a vector. More specifically, the state of a qubit is a vector in a two-dimensional vector space. This vector space is known as state space. Quantum computing relies on properties of quantum mechanics to compute problems that would be out of reach for classical computers. A quantum computer uses qubits. **Qubits** are like regular bits in a computer, but with the added ability to be put into a superposition and share entanglement with one another. Classification algorithms and methods for machine learning are essential for pattern recognition and data mining applications. Well known techniques such as support vector machines and neural networks have blossomed over the last two decades.

While still in their infancy, as quantum computers edge closer to surpassing classical computers, a new discipline is emerging called quantum machine learning. Quantum machine learning (QML) is built on two concepts: quantum data and hybrid quantum-classical models.

## II. Classical ML and Support Vector Machines

Currently, machine learning is split into three main catagories which are categories: supervised (super task-driven and is able to predict the next value), unsupervised (mostly data-driven and able to identify clusters), and reinforcement (able to learn from mistakes). We are specifically going to highlight supervised learning, which is where support vector machines fall under. Supervised learning includes using an algorithm to learn the mapping function, y=f(x) from the input to the output based on previous examples of input-output pairs.

The mapping function includes labeled training data which consists of a set of training examples. This process of the algorithm learning from the training dataset can be compared to a teacher supervising.

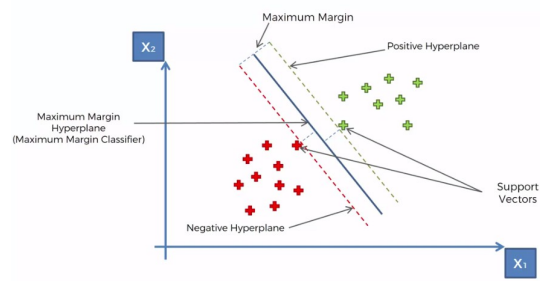

## III. Quantum ML and the Quantum SVM Algorithm

**Quantum Machine Learning (Quantum ML)** is the interdisciplinary area combining Quantum Physics and Machine Learning(ML). It is a symbiotic association- leveraging the power of Quantum Computing to produce quantum versions of ML algorithms, and applying classical ML algorithms to analyse quantum systems. Read this article for an introduction to Quantum ML. Beyond quantum computing, the term "quantum machine learning" is often associated with classical machine learning methods applied to data generated from quantum experiments (i.e. machine learning of quantum systems), such as learning quantum phase transitions or creating new quantum experiments. Quantum machine learning also extends to a branch of research that explores methodological and structural similarities between certain physical systems and learning systems, in particular neural networks.

Quantum computers will provide the computational advantage to classify objects in nth dimensions that are extremely hard to scale on classical computers. They utilise the laws of quantum mechanics, specifically superposition and entanglement to create an unthinkable amount of computational power. Quantum chips are created using that display these properties and are used to map out computer algorithms to solve complex problems. A specific problem that can be solved using quantum computers and machine learning is designing new chemical compounds. We can use ML to understand how to classify complex objects such as atoms. Except this is extremely hard to simulate on classical computers because as more atoms are added together to create larger molecules, the in-motion particles become impossible to simulate computationally as the magnitude of interactions grows too complex. In short quantum machine learning will help boost the computational power to tackle problems that are too complex to solve. Quantum Machine Learning is the analysis of classical data and quantum states on a quantum computer. The whole essence of it is being able to compute immense quantities of data more intelligently and quickly, providing the computational advantage of classifying objects too complex for classical computers and more thorough data analysis. The quantum SVM algorithm takes the classical machine learning algorithm and performs the support vector machine on a quantum circuit in order to be efficiently processed on a quantum computer.

**Support Vector Machine (SVM)** can classify objects in the nth-dimensional space where the decision based boundary is calculated using d-1. This means that the hyperplane separating your data will be one dimension lower than the feature space with all data points.

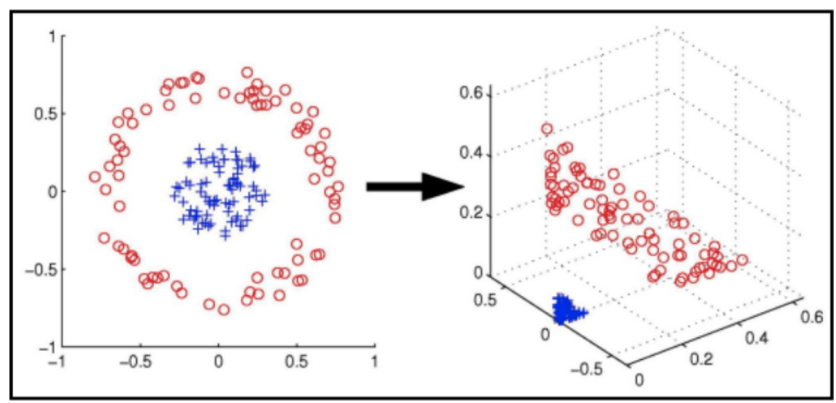

*Fig(2) The Kernel Trick. The left side is the original scenario, and the right side is the result of the mapping after the kernel trick is applied.*

# IV. Quantum Approach to Binary Classification

When data points are projected in higher dimensions, it is hard for classical computers to compute through such large computations. Hence with a quantum computer this task can become easy and even the most complex dataset can be computed in extremely high dimensions for accurate classification. This specific algorithm is called Support Vector Machine Quantum Kernel Algorithm. It is just taking the classical ML algorithm and translating it into a quantum circuit so it can be run on a quantum computer efficiently.

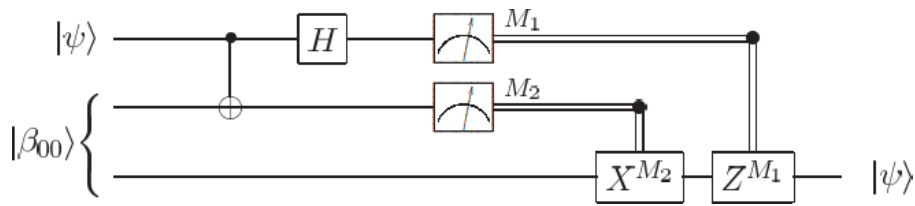

*Fig(3) An example of a quantum circuit showcasing quantum teleportation*

First step is preparing the breast-cancer dataset for the quantum circuit and importing it so the algorithm can be run. The steps are:

(1) Import dataset 'breast_cancer' or 'ad_hoc_data' or any other dataset. (2) Define your variables to include training datasets, testing dataset and how it's split. (3) Normalize dataset to have 0 unit variance so the pixels from the images have a very small range and can be computed efficiently. (4) Using IBM's qiskit library take your current features of dataset and transform them to the number of qubits. (5) Set the range for SVM to -1 and +1 so classification can be done based on where a datapoint lies on the range. (6) Set up the training dataset.
(7) Set up a plot to showcase visually the classification.

Next step is implementing the quantum ML algorithm. IBM has it part of their library which you can easily simulate. The steps are:

(1) Setting up how many qubits your quantum circuit will have. (2) Defining your classes again and import the dataset which you manipulated above. (3) Dictating the algorithm and setting its parameters for how many runs it will do and depth of the circuit. (4) Input the new datapoints (5) Check for results after the algorithm is finished running.

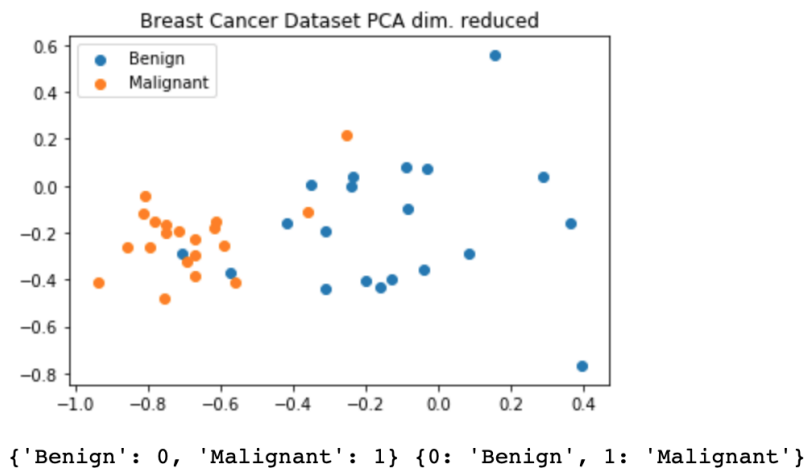

```
{'Benign': 0, 'Malignant': 1} {0: 'Benign', 1: 'Malignant'}
```

## V. Results

Based on the results, the quantum SVM is able to simulate better results to a classical SVM. Quantum Computing today is able to do many tasks in machine learning from conventional computers, and is able to solve problems that classical computers are not able to solve, given its potential. The quantum algorithm was also able to solve the problem faster in this case, having the ability to solve optimization problems in logarithmic order as opposed to the polynomic way used classically. This implementation is growing ever more efficient with the increase in the number of qubits available, and will be able to classify these large and complex datasets at a lower computational cost than what is currently available with classical computers.

## VI. Implications

Quantum Computing is still a new and emerging field, with Quantum Machine Learning even more recent. The future implications for this field are endless, reaching the realms of security, finance, medicine and so much more. An example is applying QML in examining all the possible scenarios in drug interaction and be able to present the best possible plan of action and an individual's success with that drug. Especially through a more rigorous understanding of protein folding, it would lead to more precise treatments and overall better outcomes.

## VII. Limitations

The kernel function is restricted. Linear and polynomial kernels are easy to implement, in fact, the inner product is evaluated directly in the embedding space. Radial basis function kernels are much harder to imagine in this framework, and these are the kernels the most common.

The approach only works for dense training vectors, for obvious reasons. In many classical algorithms, execution speed -- and not complexity -- is improved by exploiting sparse data structures. Such structures are common in applications, such as text mining and recommendation engines. It is unclear how a sparse vector maps to a quantum state.